\documentclass[twocolumn,showpacs,preprintnumbers,amsmath,amssymb]{revtex4}
\usepackage{amssymb}
\usepackage{graphicx}
\usepackage{dcolumn}
\usepackage{bm}
\usepackage{textcomp}

\newcommand\oprod[2]{\ensuremath{|#1\rangle\langle#2|}}

\begin{document}
\title{Generalized three-intensity decoy state method for measurement device independent quantum key distribution}
\author{Zong-Wen Yu$ ^{1,2}$, Yi-Heng Zhou$ ^{1}$,
and Xiang-Bin Wang$ ^{1,3\footnote{Email
Address: xbwang@mail.tsinghua.edu.cn}}$}

\affiliation{ \centerline{$^{1}$State Key Laboratory of Low
Dimensional Quantum Physics, Tsinghua University, Beijing 100084,
People¡¯s Republic of China}\centerline{$^{2}$Data Communication Science and Technology Research Institute, Beijing 100191, China}\centerline{$^{3}$ Shandong
Academy of Information and Communication Technology, Jinan 250101,
People¡¯s Republic of China}}

\begin{abstract}
We present improved results and more general results for the decoy-state measurement device independent quantum key distribution (MDI-QKD) where the intensities of all 3 pulses can be non-zero.  We present a more tightened explicit formula for the lower bound of the yield of two single-photon pulses for this generalized 3-intensity protocol, which can be applied to the recently proposed MDI-QKD with imperfect single-photon sources such as the coherent states or the heralded states from the parametric down-conversion. We strictly prove that our result is better than the prior art result. With this result, the final key rate rises drastically.
\end{abstract}


\maketitle


\section{Introduction}
Recently, much attention has been spent on the security of real set-ups in quantum key distribution (QKD)~\cite{BB84,GRTZ02}.  After the proposal of the decoy state method~\cite{ILM,H03,wang05,LMC05,AYKI,haya,peng,wangyang,rep,njp}, the security of QKD with an imperfect single photon source is guaranteed given whatever Eve's attack to the channel, including the famous photon-number splitting attack~\cite{PNS1,PNS}.

However, the guaranteed security assumes that Eve cannot attack device inside Alice or Bob's lab. As was demonstrated~\cite{lyderson}, given a lossy channel and limited detection efficiency, Eve can attack the gated-mode detectors inside Alice or Bob's lab if there is no additional device to detect the abnormally strong light at Alice or Bob's lab.
To reach the goal of ultimate security, one can, in principle use the device independent protocol~\cite{ind1}. However, these protocols seem to be technically demanding at the moment.

Recently,  measurement device independent QKD (MDI-QKD) was proposed based on the idea of
entanglement swapping~\cite{ind3,ind2}.
If we want to obtain a higher key rate, we can choose to directly use an imperfect single-photon source such as the coherent state~\cite{ind2} with decoy-state method for this, say the MDI decoy-state method.  Calculation formulas for the practical decoy-state implementation with only a few different states has been studied in, e.g., Refs.~\cite{wangPRA2013,qing3}, and then further studied  both experimentally~\cite{tittel1,tittel2,liuyang} and theoretically~\cite{lopa,han,curty,Wang2013}.
Theoretical study for the decoy-state MDI-QKD with only a few intensities are of particularly interesting.
The first formula for 3-intensity decoy-state MDI-QKD was proposed in~\cite{wangPRA2013}. It assumes that each user uses three different states, one vacuum state and two non-vacuum states. The results can apply to sources such as the coherent states, the parametric down conversion, and so on. It was then extended to the case that the 3 intensities are all non-vacuum coherent states~\cite{lopa}.
 One can see that the major formula there is identical with the one in Ref.\cite{wangPRA2013} in the case when Alice and Bob use coherent pulses and the weakest pulse is vacuum. In Ref.\cite{Wang2013,Wang201309}, we presented an improved explicit formulas of 3-state decoy-state method for the MDI-QKD with the weakest pulse being vacuum.

Here in this work, we shall present an improved results for the decoy-state MDI-QKD where all 3 intensities are non-vacuum. Compared with the prior-art result~\cite{lopa}, our improved formulas here make a more tightened faithful estimation of the yield and the phase flip rate of the single-photon pulse pairs and hence can raise the final key rate drastically. Also, our result is for a general type of sources with a simple condition, it does not limit to the coherent-state source. In the second section, we present the improved explicit formulas for 3-intensity decoy-state MDI-QKD and prove analytically that our result is better than the previous~\cite{lopa}. In the third section, we give the numerical simulations. The article is ended with a concluding remark.

\section{Generalized Three-intensity decoy state method for MDI-QKD}
In the protocol, each time a pulse-pair (two-pulse state) is sent to the relay for detection. The relay is controlled by an un-trusted third party (UTP). The UTP will announce whether the pulse-pair has caused a successful event. Those bits corresponding to successful events will be post-selected and further processed for the final key. Since real set-ups only use imperfect single-photon sources, we need the decoy-state method for security.

We assume Alice (Bob) has three sources, $v_A,d_A,s_A$ ($v_B,d_B,s_B$) which can only emit three different states $\rho_{v_A}, \rho_{d_A}, \rho_{s_A}$ ($\rho_{v_B}, \rho_{d_B}, \rho_{s_B}$), respectively, in photon number space. It's worth pointing out that our protocol is applicable to the weakest pulse being vacuum or not. Suppose
\begin{equation}\label{rhoAB}
  \left\{\begin{array}{lcr}
    \rho_{v_A}=\sum_{k} a^{v}_k \oprod{k}{k}, &\quad& \rho_{v_B}=\sum_{k} b^{v}_k \oprod{k}{k}, \\
    \rho_{d_A}=\sum_{k} a^{d}_k \oprod{k}{k}, &\quad& \rho_{d_B}=\sum_{k} b^{d}_k \oprod{k}{k}, \\
    \rho_{s_A}=\sum_{k} a^{s}_k \oprod{k}{k}, &\quad& \rho_{s_B}=\sum_{k} b^{s}_k \oprod{k}{k}.
  \end{array}\right.
\end{equation}

At each time, Alice will randomly select one of her 3 sources to emit a pulse, and so does Bob. The pulse form Alice and the pulse from Bob form a pulse pair and are sent to the UTP.  We regard equivalently that each time a two-pulse source is selected and a pulse pair (one pulse from Alice, one pulse from Bob) is emitted.
There are many different two-pulse sources used in the protocol. We denote $\alpha\beta$ for the two pulse source when the pulse-pair is produced by source
$\alpha$ at Alice's side and source $\beta$ at Bob's side, $\alpha$ can be one of $\{v_A,d_A,s_A\}$ and $\beta$
can be one of $\{v_B,d_B,s_B\}$. For example, at a certain time $j$ Alice uses source $v_A$ and Bob uses source $s_B$, we say the pulse pair is emitted by source $v_As_B$.

In the protocol, two different bases (e.g., $Z$ basis consisting of horizontal polarization $|H\rangle\langle H|$  and vertical polarization $|V\rangle\langle V|$, and $X$ basis consisting of $\pi/4$ and $3\pi/4$ polarizations) are used. The density operator in photon number space alone does not describe the state in the composite space. We shall apply the decoy-state method analysis in the same basis (e.g., $Z$ basis or $X$ basis) for pulses from sources $\alpha\beta(\alpha\in\{v_A,d_A,s_A\},\beta\in\{v_B,d_B,s_B\})$. Therefore we only need consider the density operators in the photon number space.

According to the decoy-state theory, the yield of a certain set of pulse pairs is defined as the happening rate of a successful event (announced by the UTP) corresponding to pulse pairs out of the set. We use capital letter $Y_{\alpha\beta}$ to denote the nine yields for sources $\alpha\beta(\alpha\in\{v_A,d_A,s_A\},\beta\in\{v_B,d_B,s_B\})$ and these values can be directly calculated from the observed experimental data. We use $y_{mn}$ to denotes the yield given that Alice and Bob send respectively an $m$-photon and $n$-photon pulses.

Our first major task is to deduce $y_{11}$ from the known values, i.e., to formulate $y_{11}$, the yield of state $\oprod{1}{1}\otimes \oprod{1}{1}$ in the nine known values $\{Y_{\alpha\beta}|\alpha\in\{v_A,d_A,s_A\},\beta\in\{v_B,d_B,s_B\}\}$. Using the convex proposition, we have the following relations
\begin{equation}\label{Yalphabeta}
  Y_{\alpha\beta}=\sum_{k,l\geq 0}a^{\alpha}_{k}b^{\beta}_{l}y_{kl},
\end{equation}
where $\alpha\in\{v_A,d_A,s_A\}$ and $\beta\in\{v_B,d_B,s_B\}$. In practice, the imperfect single-photon sources have the property that the probability of emitting an empty pulse is greater than the probability of the no-empty pulse. With this fact, in order to estimate the lower bound of $y_{11}$, we need to eliminate the unknown values $y_{0l}(l\geq 0)$ and $y_{k0}(k\geq 0)$ firstly.

Without causing any ambiguity, we omit the subscripts A and B in the following of this paper. Cancel out the terms $y_{0k}$ and $y_{j0}$ using Gaussian elimination, we obtain the following four relations
\begin{eqnarray}\label{hatY}
 \left\{\begin{array}{c}
  a_0^{v}b_0^v \hat{Y}_{dd}-a_0^v b_0^d\hat{Y}_{dv}-a_0^d b_0^v \hat{Y}_{vd}+a_0^d b_0^d\hat{Y}_{vv}=0 ,\\
  a_0^{v}b_0^v \hat{Y}_{ds}-a_0^v b_0^s\hat{Y}_{dv}-a_0^d b_0^v \hat{Y}_{vs}+a_0^d b_0^s\hat{Y}_{vv}=0 ,\\
  a_0^{v}b_0^v \hat{Y}_{sd}-a_0^v b_0^d\hat{Y}_{sv}-a_0^s b_0^v \hat{Y}_{vd}+a_0^s b_0^d\hat{Y}_{vv}=0 ,\\
  a_0^{v}b_0^v \hat{Y}_{ss}-a_0^v b_0^s\hat{Y}_{sv}-a_0^s b_0^v \hat{Y}_{vs}+a_0^s b_0^s\hat{Y}_{vv}=0 ,
 \end{array}\right.
\end{eqnarray}
where $\hat{Y}_{ij}=Y_{ij}-\sum_{k,l\geq 1}a_{k}^{i}b_{l}^{j}y_{kl}, (i,j\in\{v,d,s\})$. For the sake of convenience, we denote
\begin{eqnarray}\label{tildeY}
 \left\{\begin{array}{c}
  \tilde{Y}_{dd}^{\prime}=a_0^{v}b_0^v {Y}_{dd}-a_0^v b_0^d {Y}_{dv}-a_0^d b_0^v {Y}_{vd}+a_0^d b_0^d {Y}_{vv},\\
  \tilde{Y}_{ds}^{\prime}=a_0^{v}b_0^v {Y}_{ds}-a_0^v b_0^s {Y}_{dv}-a_0^d b_0^v {Y}_{vs}+a_0^d b_0^s {Y}_{vv},\\
  \tilde{Y}_{sd}^{\prime}=a_0^{v}b_0^v {Y}_{sd}-a_0^v b_0^d {Y}_{sv}-a_0^s b_0^v {Y}_{vd}+a_0^s b_0^d {Y}_{vv},\\
  \tilde{Y}_{ss}^{\prime}=a_0^{v}b_0^v {Y}_{ss}-a_0^v b_0^s {Y}_{sv}-a_0^s b_0^v {Y}_{vs}+a_0^s b_0^s {Y}_{vv},
 \end{array}\right.
\end{eqnarray}
and introduce the following notations
\begin{equation}\label{defh}
  h_{x_k}^{i}=\frac{x_k^i}{x_0^i},\quad (k\geq 0;x\in\{a,b\};i\in\{v,d,s\}).
\end{equation}
Divide the four relations in Eq.(\ref{hatY}) by the positive factors $a_0^v a_0^d b_0^v b_0^d$, $a_0^v a_0^d b_0^v b_0^s$, $a_0^v a_0^s b_0^v b_0^d$ and $a_0^v a_0^s b_0^v b_0^s$ respectively, we obtain the following simplified equation
\begin{eqnarray}\label{tildeY11}
 \left\{\begin{array}{c}
  \tilde{Y}_{dd}=\sum_{k,l\geq 1}(h_{a_k}^d-h_{a_k}^v)(h_{b_l}^d-h_{b_l}^v)y_{kl},\\
  \tilde{Y}_{ds}=\sum_{k,l\geq 1}(h_{a_k}^d-h_{a_k}^v)(h_{b_l}^s-h_{b_l}^v)y_{kl},\\
  \tilde{Y}_{sd}=\sum_{k,l\geq 1}(h_{a_k}^s-h_{a_k}^v)(h_{b_l}^d-h_{b_l}^v)y_{kl},\\
  \tilde{Y}_{ss}=\sum_{k,l\geq 1}(h_{a_k}^s-h_{a_k}^v)(h_{b_l}^s-h_{b_l}^v)y_{kl}.
 \end{array}\right.
\end{eqnarray}
where $\tilde{Y}_{ij}$ are defined as follows
\begin{equation}
  \tilde{Y}_{ij}=\frac{\tilde{Y}_{ij}^{\prime}}{a_0^v a_0^i b_0^v b_0^j}, \quad (i,j\in\{d,s\}),
\end{equation}
with $\tilde{Y}_{ij}^{\prime}$ being presented in Eq.(\ref{tildeY}). Furthermore, if we introduce
\begin{equation}\label{defht}
  \tilde{h}_{x_k}^{i}=h_{x_k}^{i}-h_{x_k}^{v}, \quad (k\geq 0; x\in\{a,b\}; i\in \{v,d,s\}),
\end{equation}
with $h_{x_k}^{i}$ defined in Eq.(\ref{defh}), we can write the relations about variables $y_{kl}$ given by Eq.(\ref{tildeY11}) into the following simplified form
\begin{eqnarray}
  \tilde{Y}_{dd}&=&\tilde{h}_{a_1}^d\tilde{h}_{b_1}^d y_{11}+ \tilde{h}_{a_1}^d\tilde{h}_{b_2}^d y_{12} + \tilde{h}_{a_2}^d\tilde{h}_{b_1}^d y_{21}+ J_{dd}, \label{tYdd} \\
  \tilde{Y}_{ds}&=&\tilde{h}_{a_1}^d\tilde{h}_{b_1}^s y_{11}+ \tilde{h}_{a_1}^d\tilde{h}_{b_2}^s y_{12} + \tilde{h}_{a_2}^d\tilde{h}_{b_1}^s y_{21}+ J_{ds}, \label{tYds} \\
  \tilde{Y}_{sd}&=&\tilde{h}_{a_1}^s\tilde{h}_{b_1}^d y_{11}+ \tilde{h}_{a_1}^s\tilde{h}_{b_2}^d y_{12} + \tilde{h}_{a_2}^s\tilde{h}_{b_1}^d y_{21}+ J_{sd}, \label{tYsd} \\
  \tilde{Y}_{ss}&=&\tilde{h}_{a_1}^s\tilde{h}_{b_1}^s y_{11}+ \tilde{h}_{a_1}^s\tilde{h}_{b_2}^s y_{12} + \tilde{h}_{a_2}^s\tilde{h}_{b_1}^s y_{21}+ J_{ss}, \label{tYss}
\end{eqnarray}
where $\tilde{Y}_{ij}$ is defined in Eq.(\ref{tildeY}), $\tilde{h}_{x_k}^{i}$ is given by Eq.(\ref{defht}), and $J_{ij}=\sum_{(k,l)\in J_1}{\tilde{h}_{a_k}^{i}\tilde{h}_{b_l}^{j} y_{kl}},(i,j\in\{d,s\})$ with $J_1=\{(k,l)|k,l\geq 1; k+l\geq 4\}$.

With these preparations, now we are going to introduce the following very important condition:
\begin{equation}\label{cond1}
  \frac{\tilde{h}_{x_k}^s}{\tilde{h}_{x_k}^d}\geq \frac{\tilde{h}_{x_2}^s}{\tilde{h}_{x_2}^d} \geq \frac{\tilde{h}_{x_1}^s}{\tilde{h}_{x_1}^d}, \quad (x\in\{a,b\})
\end{equation}
for $k\geq 2$. We can easily prove that the imperfect sources used in practice such as the coherent state source, the heralded source out of the parametric-down conversion, satisfy the above restriction.

In order to get a lower bound of $y_{11}$, we should derive the expression of $y_{11}$ with Eqs.(\ref{tYdd}-\ref{tYss}) firstly. Combining Eqs.(\ref{tYdd}-\ref{tYsd}), we obtain the expression of $y_{11}$ by eliminating $y_{12}$ and $y_{21}$ such that
\begin{equation}\label{y11with123}
  y_{11}=y_{11}^{(123)}+\sum_{(m,n)\in J_{1}}f_{11}^{(123)}(m,n)y_{mn},
\end{equation}
where $J_{1}=\{(m,n)|m\geq 1, n\geq 1,m+n\geq 4\}$,
\begin{widetext}
\begin{equation}\label{y11of123}
  y_{11}^{(123)}=\frac{(\tilde{h}_{a_1}^d \tilde{h}_{a_2}^s \tilde{h}_{b_1}^d \tilde{h}_{b_2}^s-\tilde{h}_{a_1}^s \tilde{h}_{a_2}^d \tilde{h}_{b_1}^s \tilde{h}_{b_2}^d)\tilde{Y}_{dd}-\tilde{h}_{b_1}^d \tilde{h}_{b_2}^d (\tilde{h}_{a_1}^d \tilde{h}_{a_2}^{s}-\tilde{h}_{a_1}^s \tilde{h}_{a_2}^d)\tilde{Y}_{ds}-\tilde{h}_{a_1}^{d} \tilde{h}_{a_2}^d (\tilde{h}_{b_1}^d \tilde{h}_{b_2}^s-\tilde{h}_{b_1}^s \tilde{h}_{b_2}^d)\tilde{Y}_{sd}} {\tilde{h}_{a_1}^d \tilde{h}_{b_1}^d(\tilde{h}_{a_1}^d \tilde{h}_{a_2}^s-\tilde{h}_{a_1}^s \tilde{h}_{a_2}^d) (\tilde{h}_{b_1}^d \tilde{h}_{b_2}^s-\tilde{h}_{b_1}^s \tilde{h}_{b_2}^d)},
\end{equation}
and
\begin{equation}\label{f11of123}
  f_{11}^{(123)}(m,n)=\frac{\tilde{h}_{a_2}^d \tilde{h}_{b_n}^d(\tilde{h}_{a_1}^d \tilde{h}_{a_m}^s-\tilde{h}_{a_1}^s \tilde{h}_{a_m}^d)(\tilde{h}_{b_1}^d \tilde{h}_{b_2}^s-\tilde{h}_{b_1}^s \tilde{h}_{b_2}^d)+\tilde{h}_{a_m}^d \tilde{h}_{b_1}^d(\tilde{h}_{a_1}^d \tilde{h}_{a_2}^s-\tilde{h}_{a_1}^s \tilde{h}_{a_2}^d)(\tilde{h}_{b_2}^d \tilde{h}_{b_n}^s-\tilde{h}_{b_2}^s \tilde{h}_{b_n}^d)}{\tilde{h}_{a_1}^d \tilde{h}_{b_1}^d(\tilde{h}_{a_1}^d \tilde{h}_{a_2}^s-\tilde{h}_{a_1}^s \tilde{h}_{a_2}^d) (\tilde{h}_{b_1}^d \tilde{h}_{b_2}^s-\tilde{h}_{b_1}^s \tilde{h}_{b_2}^d)}.
\end{equation}
\end{widetext}
In these expressions, we use the superscript $*^{(123)}$ to denote the result obtained with the first three equations from Eqs.(\ref{tYdd}-\ref{tYss}). Under the conditions presented in Eq.(\ref{cond1}), we can easily find out that $(\tilde{h}_{a_1}^d \tilde{h}_{a_2}^s-\tilde{h}_{a_1}^s \tilde{h}_{a_2}^d)\geq 0$, $(\tilde{h}_{b_1}^d \tilde{h}_{b_2}^s-\tilde{h}_{b_1}^s \tilde{h}_{b_2}^d)\geq 0$, $(\tilde{h}_{a_1}^d \tilde{h}_{a_m}^s-\tilde{h}_{a_1}^s \tilde{h}_{a_m}^d)\geq 0$ for all $m\geq 1$ and $(\tilde{h}_{b_2}^d \tilde{h}_{b_n}^s-\tilde{h}_{b_2}^s \tilde{h}_{b_n}^d)\geq 0$ for all $n\geq 2$. Then we know that $f_{11}^{(123)}(m,n)\geq 0$ hold for all $(m,n)\in J_1$. With this fact, we obtain a lower bound from Eq.(\ref{y11with123}) by setting $y_{mn}=0,(m,n)\in J_1$ such that
\begin{equation}\label{y11L123}
  \underline{y_{11}}=y_{11}^{(123)}\leq y_{11},
\end{equation}
where $y_{11}^{(123)}$ is defined by Eq.(\ref{y11of123}). This and Eq.(\ref{y11of123}) are our major formulas for the decoy-state method implementation for MDI-QKD in this section.

Similarly, we can get other expressions with choosing any other three equations from Eqs.(\ref{tYdd}-\ref{tYss}). For example, we choose Eqs.(\ref{tYdd}-\ref{tYds},\ref{tYss}). By eliminating $y_{12}$ and $y_{21}$, we get another expression of $y_{11}$ such that
\begin{equation}\label{y11with124}
  y_{11}= y_{11}^{(124)}+\sum_{(m,n)\in J_1} f_{11}^{(124)}(m,n)s_{mn},
\end{equation}
where
\begin{widetext}
\begin{equation}\label{y11of124}
  y_{11}^{(124)}=\frac{\tilde{h}_{b_1}^s \tilde{h}_{b_2}^s(\tilde{h}_{a_1}^d \tilde{h}_{a_2}^s-\tilde{h}_{a_1}^s \tilde{h}_{a_2}^d)\tilde{Y}_{dd}+(\tilde{h}_{a_1}^s \tilde{h}_{a_2}^d \tilde{h}_{b_1}^d \tilde{h}_{b_2}^s-\tilde{h}_{a_1}^d \tilde{h}_{a_2}^s \tilde{h}_{b_1}^s \tilde{h}_{b_2}^d)\tilde{Y}_{ds}-\tilde{h}_{a_1}^d \tilde{h}_{a_2}^d (\tilde{h}_{b_1}^d \tilde{h}_{b_2}^s-\tilde{h}_{b_1}^s \tilde{h}_{b_2}^d) \tilde{Y}_{ss}}{\tilde{h}_{a_1}^d \tilde{h}_{b_1}^s(\tilde{h}_{a_1}^d \tilde{h}_{a_2}^s-\tilde{h}_{a_1}^s \tilde{h}_{a_2}^d) (\tilde{h}_{b_1}^d \tilde{h}_{b_2}^s-\tilde{h}_{b_1}^s \tilde{h}_{b_2}^d)},
\end{equation}
and
\begin{equation}\label{f11of124}
  f_{11}^{(124)}(m,n)=\frac{\tilde{h}_{a_2}^d \tilde{h}_{b_n}^s(\tilde{h}_{a_1}^d \tilde{h}_{a_m}^s-\tilde{h}_{a_1}^s \tilde{h}_{a_m}^d)(\tilde{h}_{b_1}^d \tilde{h}_{b_2}^s-\tilde{h}_{b_1}^s \tilde{h}_{b_2}^d)+\tilde{h}_{a_m}^d \tilde{h}_{b_1}^s(\tilde{h}_{a_1}^d \tilde{h}_{a_2}^s-\tilde{h}_{a_1}^s \tilde{h}_{a_2}^d)(\tilde{h}_{b_2}^d \tilde{h}_{b_n}^s-\tilde{h}_{b_2}^s \tilde{h}_{b_n}^d)}{\tilde{h}_{a_1}^d \tilde{h}_{b_1}^s(\tilde{h}_{a_1}^d \tilde{h}_{a_2}^s-\tilde{h}_{a_1}^s \tilde{h}_{a_2}^d) (\tilde{h}_{b_1}^d \tilde{h}_{b_2}^s-\tilde{h}_{b_1}^s \tilde{h}_{b_2}^d)}.
\end{equation}
\end{widetext}
Under the conditions presented in Eq.(\ref{cond1}), we can also find out that $f_{11}^{(124)}(m,n)\geq 0$ for all $(m,n)\in J_1$. Then we know that $y_{11}^{(124)}$ is alos a lower bound of $y_{11}$. On the other hand, by comparing $f_{11}^{(123)}(m,n)$ and $f_{11}^{(124)}(m,n)$, we have
\begin{eqnarray*}\label{dif123and124}
  & &f_{11}^{(123)}(m,n)-f_{11}^{(124)}(m,n) \nonumber \\
  &=&\frac{\tilde{h}_{a_2}^d(\tilde{h}_{a_1}^d \tilde{h}_{a_m}^s-\tilde{h}_{a_1}^s \tilde{h}_{a_m}^d)(\tilde{h}_{b_1}^d \tilde{h}_{b_n}^s-\tilde{h}_{b_1}^s \tilde{h}_{b_n}^d)}{-\tilde{h}_{a_1}^d \tilde{h}_{b_1}^d \tilde{h}_{b_1}^s (\tilde{h}_{a_1}^d \tilde{h}_{a_2}^s-\tilde{h}_{a_1}^s \tilde{h}_{a_2}^d)}\leq 0,
\end{eqnarray*}
for all $(m,n)\in J_1$. Then we know that
\begin{equation}\label{y11of123and124}
  y_{11}^{(123)}\geq y_{11}^{(124)},
\end{equation}
with Eq.(\ref{y11with123}) and Eq.(\ref{y11with124}). With the relation presented in Eq.(\ref{y11of123and124}) we know that the lower bound $y_{11}^{(123)}$ is tighter than the lower bound $y_{11}^{(124)}$. In the same way, we can get another two lower bounds $y_{11}^{(134)}$ and $y_{11}^{(234)}$ of $y_{11}$ with Eqs.(\ref{tYdd},\ref{tYsd}-\ref{tYss}) and Eqs.(\ref{tYds}-\ref{tYss}) respectively. Furthermore, we can also prove that
\begin{equation}\label{dif34}
  y_{11}^{(123)}\geq y_{11}^{(134)}, \quad y_{11}^{(123)}\geq y_{11}^{(234)}.
\end{equation}

Now we only consider Eq.(\ref{tYdd}) and Eq.(\ref{tYss}). By eliminating $y_{12}$ or $y_{21}$ respectively, we get two expressions of $y_{11}$ such that
\begin{eqnarray}
  y_{11}&=&y_{11}^{(14a)}+\sum_{(m,n)\in J_{2}} f_{11}^{(14a)}(m,n) y_{mn}, \label{y11Ka} \\
  y_{11}&=&y_{11}^{(14b)}+\sum_{(m,n)\in J_{2}} f_{11}^{(14a)}(m,n) y_{mn}, \label{y11Kb}
\end{eqnarray}
where $J_2=\{(m,n)|m\geq 1,n\geq 1, m+n\geq 3\}$,
\begin{eqnarray}
  y_{11}^{(14a)}&=&\frac{\tilde{h}_{a_1}^s \tilde{h}_{b_2}^s \tilde{Y}_{dd}-\tilde{h}_{a_1}^d \tilde{h}_{b_2}^d \tilde{Y}_{ss}}{\tilde{h}_{a_1}^d \tilde{h}_{a_1}^s(\tilde{h}_{b_1}^d \tilde{h}_{b_2}^s-\tilde{h}_{b_1}^s \tilde{h}_{b_2}^d)}, \label{y11of14a} \\
  y_{11}^{(14b)}&=&\frac{\tilde{h}_{a_2}^s \tilde{h}_{b_1}^s \tilde{Y}_{dd}-\tilde{h}_{a_2}^d \tilde{h}_{b_1}^d \tilde{Y}_{ss}}{\tilde{h}_{b_1}^d \tilde{h}_{b_1}^s(\tilde{h}_{a_1}^d \tilde{h}_{a_2}^s-\tilde{h}_{a_1}^s \tilde{h}_{a_2}^d)}, \label{y11of14b}
\end{eqnarray}
and
\begin{eqnarray}
  f_{11}^{(14a)}(m,n)&=&\frac{\tilde{h}_{a_1}^d \tilde{h}_{b_2}^d \tilde{h}_{a_m}^s \tilde{h}_{b_n}^s-\tilde{h}_{a_1}^s \tilde{h}_{b_2}^s \tilde{h}_{a_m}^d \tilde{h}_{b_n}^d}{\tilde{h}_{a_1}^d \tilde{h}_{a_1}^s(\tilde{h}_{b_1}^d \tilde{h}_{b_2}^s-\tilde{h}_{b_1}^s \tilde{h}_{b_2}^d)}, \label{f11of14a} \\
  f_{11}^{(14b)}(m,n)&=&\frac{\tilde{h}_{a_2}^d \tilde{h}_{b_1}^d \tilde{h}_{a_m}^s \tilde{h}_{b_n}^s-\tilde{h}_{a_2}^s \tilde{h}_{b_1}^s \tilde{h}_{a_m}^d \tilde{h}_{b_n}^d}{\tilde{h}_{b_1}^d \tilde{h}_{b_1}^s(\tilde{h}_{a_1}^d \tilde{h}_{a_2}^s-\tilde{h}_{a_1}^s \tilde{h}_{a_2}^d)}. \label{f11of14b}
\end{eqnarray}
For any sources used in the protocol, we must have either
\begin{equation*}
  K_a=\frac{\tilde{h}_{a_1}^s \tilde{h}_{b_2}^s}{\tilde{h}_{a_1}^d \tilde{h}_{b_2}^d}\leq \frac{\tilde{h}_{a_2}^s \tilde{h}_{b_1}^s}{\tilde{h}_{a_2}^d \tilde{h}_{b_1}^d}=K_b \quad \textrm{or}\quad  K_a> K_b.
\end{equation*}
Suppose the former one holds, we can easily find out that $f_{11}^{(14a)}(m,n)\geq 0$ for all $(m,n)\in J_2$ and $y_{11}^{(14a)}$ is a lower bound of $y_{11}$. On the other hand, if $K_a\geq K_b$ holds, we have $f_{11}^{(14b)}(m,n)\geq 0$ for all $(m,n)\in J_2$ and $y_{11}^{(14b)}$ is a lower bound of $y_{11}$. Considering the following two relations
\begin{equation}\label{kaKb}
  K_a-K_b =\frac{\tilde{h}_{a_1}^s \tilde{h}_{a_2}^d \tilde{h}_{b_1}^d \tilde{h}_{b_2}^s-\tilde{h}_{a_1}^d \tilde{h}_{a_2}^s \tilde{h}_{b_1}^s \tilde{h}_{b_2}^d}{\tilde{h}_{a_1}^d \tilde{h}_{a_2}^d \tilde{h}_{b_1}^d \tilde{h}_{b_2}^d},
\end{equation}
and
\begin{eqnarray*}\label{f11of14ab}
  & &f_{11}^{(14a)}(m,n)-f_{11}^{(14b)}(m,n) \nonumber \\
  &=&\frac{\tilde{h}_{a_2}^d \tilde{h}_{b_2}^d(\tilde{h}_{a_1}^d \tilde{h}_{a_m}^s \tilde{h}_{b_1}^d \tilde{h}_{b_n}^s-\tilde{h}_{a_1}^s \tilde{h}_{a_m}^d \tilde{h}_{b_1}^s \tilde{h}_{b_n}^d)(K_a-K_b)}{\tilde{h}_{a_1}^s \tilde{h}_{b_1}^s (\tilde{h}_{a_1}^d \tilde{h}_{a_2}^s-\tilde{h}_{a_1}^s \tilde{h}_{a_2}) (\tilde{h}_{b_1}^d \tilde{h}_{b_2}^s-\tilde{h}_{b_1}^s \tilde{h}_{b_2}^d)},
\end{eqnarray*}
we know that $K_a-K_b$ and $f_{11}^{(14a)}-f_{11}^{(14b)}$ have the same sign which means that they are both positive or negative simultaneously. Then we can write the lower bound of $y_{11}$ with Eq.(\ref{tYdd}) and Eq.(\ref{tYss}) into the following compact form
\begin{equation}\label{y11of14}
  y_{11}^{(14)}=\min\{y_{11}^{(14a)},y_{11}^{(14b)}\}.
\end{equation}
If both Alice and Bob use coherent pulses, the lower bound given by Eq.(\ref{y11of14}) is just the result presented in Refs.~\cite{lopa,curty}. In the coming, we will prove that the lower bound $y_{11}^{(123)}$ given in Eq.(\ref{y11of123}) is more tightly than $y_{11}^{(14)}$. Firstly, if we suppose $K_a\leq K_b$ holds, then we know that $\tilde{h}_{a_1}^s \tilde{h}_{a_2}^d \tilde{h}_{b_1}^d \tilde{h}_{b_2}^s\leq \tilde{h}_{a_1}^d \tilde{h}_{a_2}^s \tilde{h}_{b_1}^s \tilde{h}_{b_2}^d$ and $y_{11}^{(14)}=y_{11}^{(14a)}$. For any $(m,n)\in J_1$ we have
\begin{eqnarray*}
  & &f_{11}^{(123)}(m,n)-f_{11}^{(14a)}(m,n) \nonumber \\
  &=&-\frac{(\tilde{h}_{a_1}^d \tilde{h}_{a_m}^s-\tilde{h}_{a_1}^s \tilde{h}_{a_m})D_a}{\tilde{h}_{a_1}^d \tilde{h}_{a_1}^s \tilde{h}_{b_1}^d (\tilde{h}_{a_1}^d \tilde{h}_{a_2}^s-\tilde{h}_{a_1}^s \tilde{h}_{a_2}) (\tilde{h}_{b_1}^d \tilde{h}_{b_2}^s-\tilde{h}_{b_1}^s \tilde{h}_{b_2}^d)},
\end{eqnarray*}
where $D_a=(\tilde{h}_{a_1}^d \tilde{h}_{a_2}^s \tilde{h}_{b_1}^d \tilde{h}_{b_2}^d \tilde{h}_{b_n}^s
+ \tilde{h}_{a_1}^s \tilde{h}_{a_2}^d \tilde{h}_{b_1}^s \tilde{h}_{b_2}^d \tilde{h}_{b_n}^d
+ \tilde{h}_{a_1}^s \tilde{h}_{a_2}^d \tilde{h}_{b_1}^d \tilde{h}_{b_2}^s \tilde{h}_{b_n}^d
+ \tilde{h}_{a_1}^s \tilde{h}_{a_2}^d \tilde{h}_{b_1}^d \tilde{h}_{b_2}^d \tilde{h}_{b_n}^s)\geq \tilde{h}_{b_2}^d (\tilde{h}_{a_1}^d \tilde{h}_{a_2}^s-\tilde{h}_{a_1}^s \tilde{h}_{a_2}^d)(\tilde{h}_{b_1}^d \tilde{h}_{b_n}^s-\tilde{h}_{b_1}^s \tilde{h}_{b_n}^d)$. Then we know that
\begin{eqnarray*}
  & &f_{11}^{(123)}(m,n)-f_{11}^{(14a)}(m,n) \nonumber \\
  &\leq & -\frac{\tilde{h}_{b_2}^d (\tilde{h}_{a_1}^d \tilde{h}_{a_m}^s-\tilde{h}_{a_1}^s \tilde{h}_{a_m}^d)(\tilde{h}_{b_1}^d \tilde{h}_{b_n}^s-\tilde{h}_{b_1}^s \tilde{h}_{b_n}^d)}{\tilde{h}_{a_1}^d \tilde{h}_{a_1}^s \tilde{h}_{b_1}^d (\tilde{h}_{b_1}^d \tilde{h}_{b_2}^s-\tilde{h}_{b_1}^s \tilde{h}_{b_2}^d)} \leq 0.
\end{eqnarray*}
We can easily know that $y_{11}^{(123)}\geq y_{11}^{(14)}$ when $K_a\leq K_b$ with this equation. Secondly, if we suppose $K_a> K_b$ holds, we can easily prove that $f_{11}^{(123)}(m,n)-f_{11}^{(14b)}(m,n)\leq 0$ for all $(m,n)\in J_1$ within the same way. Then we get $y_{11}^{(123)}\geq y_{11}^{(14)}$ when $K_a> K_b$. This completes the proof that $y_{11}^{(123)}\geq y_{11}^{(14)}$.

In order to estimate the final key rate, we also need the upper bound of error rate caused by the two single-photon pulses, say $e_{11}$. Similar to the total gain, the total error rate with source $\alpha\beta$ chosen by Alice and Bob can be written as~\cite{ind2}
\begin{eqnarray}
  \tilde{T}_{dd}&=&\tilde{h}_{a_1}^d\tilde{h}_{b_1}^d t_{11}+ \tilde{h}_{a_1}^d\tilde{h}_{b_2}^d t_{12} + \tilde{h}_{a_2}^d\tilde{h}_{b_1}^d t_{21}+ K_{dd}, \label{Tdd} \\
  \tilde{T}_{ds}&=&\tilde{h}_{a_1}^d\tilde{h}_{b_1}^s t_{11}+ \tilde{h}_{a_1}^d\tilde{h}_{b_2}^s t_{12} + \tilde{h}_{a_2}^d\tilde{h}_{b_1}^s t_{21}+ K_{ds}, \label{Tds} \\
  \tilde{T}_{sd}&=&\tilde{h}_{a_1}^s\tilde{h}_{b_1}^d t_{11}+ \tilde{h}_{a_1}^s\tilde{h}_{b_2}^d t_{12} + \tilde{h}_{a_2}^s\tilde{h}_{b_1}^d t_{21}+ K_{sd}, \label{Tsd} \\
  \tilde{T}_{ss}&=&\tilde{h}_{a_1}^s\tilde{h}_{b_1}^s t_{11}+ \tilde{h}_{a_1}^s\tilde{h}_{b_2}^s t_{12} + \tilde{h}_{a_2}^s\tilde{h}_{b_1}^s t_{21}+ K_{ss}, \label{Tss}
\end{eqnarray}
where $T_{ij}=Y_{ij}E_{ij},(i,j\in\{v,d,s\})$, $t_{mn}=y_{mn}e_{mn}$, and
\begin{eqnarray}
  \tilde{T}_{dd}&=&\frac{T_{dd}}{a_0^d b_0^d}-\frac{T_{dv}}{a_0^d b_0^v}-\frac{T_{vd}}{a_0^v b_0^d}+\frac{T_{vv}}{a_0^v b_0^v}, \label{tTxx} \\
  \tilde{T}_{ds}&=&\frac{T_{ds}}{a_0^d b_0^s}-\frac{T_{dv}}{a_0^d b_0^v}-\frac{T_{vs}}{a_0^v b_0^s}+\frac{T_{vv}}{a_0^v b_0^v}, \label{tTxy} \\
  \tilde{T}_{sd}&=&\frac{T_{sd}}{a_0^s b_0^d}-\frac{T_{sv}}{a_0^s b_0^v}-\frac{T_{vd}}{a_0^v b_0^d}+\frac{T_{vv}}{a_0^v b_0^v}, \label{tTyx} \\
  \tilde{T}_{ss}&=&\frac{T_{ss}}{a_0^s b_0^s}-\frac{T_{sv}}{a_0^s b_0^v}-\frac{T_{vs}}{a_0^v b_0^s}+\frac{T_{vv}}{a_0^v b_0^v}, \label{tTyy}
\end{eqnarray}
and $K_{ij}=\sum_{(k,l)\in J_1}{\tilde{h}_{a_k}^{i}\tilde{h}_{b_l}^{j} t_{kl}},(i,j\in\{d,s\})$ with $J_1=\{(k,l)|k,l\geq 1,k+l\geq 4\}$. According to Eq.(\ref{Tdd}), we can find out the upper bound of $e_{11}$ such that
\begin{equation}\label{e11U}
  e_{11}\leq e_{11}^{(1)}=\frac{\tilde{T}_{dd}}{\tilde{h}_{a_1}^{d} \tilde{h}_{b_1}^d y_{11}}=\overline{e_{11}}.
\end{equation}

In the protocol, there are two different bases. We denote $y_{11}^{Z}$ and $y_{11}^{X}$ for yields of single-photon pulse pairs in the $Z$ and $X$ bases, respectively. Consider those post-selected bits cased by source $d_A d_B$ in the $Z$ basis. After an error test, we know the bit-flip error rate of this set, say $T_{d_A d_B}^{Z}=Y_{d_A d_B}^{Z}E_{d_A d_B}^{Z}$. We also need the phase-flip rate for the subset of bits which are caused by the two single-photon pulse, say $e_{11}^{ph}$, which is equal to the flip rate of post-selected bits caused by a single photon in the $X$ basis, say $e_{11}^{X}$. Given this, we can now calculate the key rate by the well-know formula. For example, for those post-selected bits caused by source $s_{A} s_{B}$, it is
\begin{equation}\label{KeyRate}
  R=a_1^s b_1^s y_{11}^{Z}[1-H(e_{11}^{X})]-Y_{s_A s_B}^{Z}fH(E_{s_A s_B}^{Z}),
\end{equation}
where $f$ is the efficiency factor of the error correction method used.

\section{Numerical Simulation}

\begin{figure}
  \includegraphics[width=240pt]{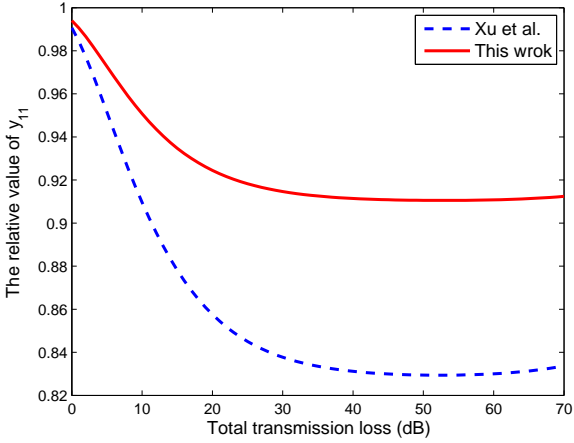}\\
  \caption{(Color online) The relative value between estimated parameter of $y_{11}$ and the asymptotic limit of the infinite decoy-state method versus the total channel transmission loss using 3-intensity decoy state MDI-QKD. In the simulations, we assume that $\mu_v=\nu_v=0.01$, $\mu_d=\nu_d=0.1$,  $\mu_s=\nu_s=0.5$ and the value of other parameters are presented in Table~\ref{tabPara}.}\label{ry11Lp1p5}
\end{figure}

\begin{figure}
  \includegraphics[width=240pt]{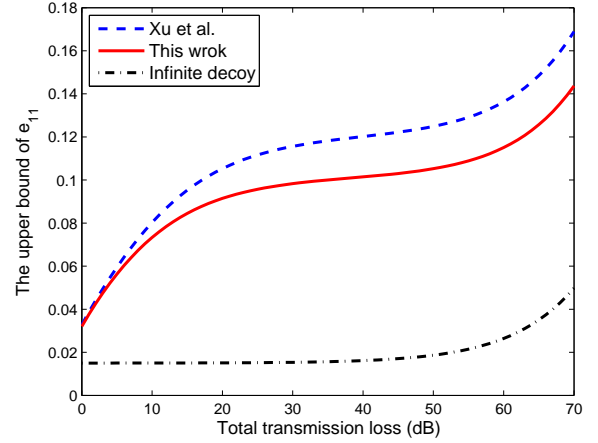}\\
  \caption{(Color online) The estimated parameter of $e_{11}$ versus the total channel transmission loss using 3-intensity decoy state MDI-QKD. In the simulations, we assume that $\mu_v=\nu_v=0.01$, $\mu_d=\nu_d=0.1$,  $\mu_s=\nu_s=0.5$ and the value of other parameters are presented in Table~\ref{tabPara}.}\label{e11Up1p5}
\end{figure}

\begin{figure}
  \includegraphics[width=240pt]{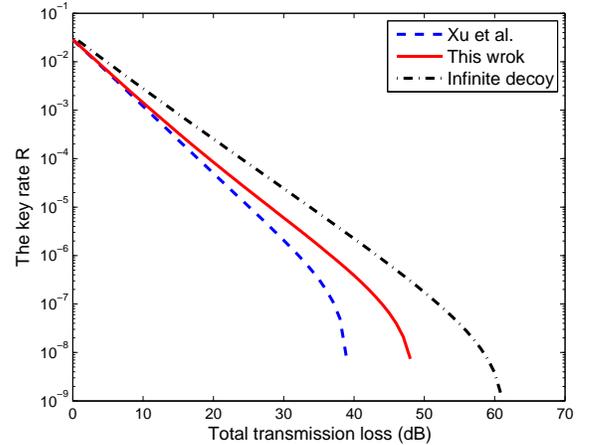}\\
  \caption{(Color online) The estimated key rate $R$ versus channel transmission using 3-intensity decoy state MDI-QKD. In the simulations, we assume that $\mu_v=\nu_v=0.01$, $\mu_d=\nu_d=0.1$,  $\mu_s=\nu_s=0.5$ and the value of other parameters are presented in Table~\ref{tabPara}.}\label{Rp1p5}
\end{figure}

\begin{figure}
  \includegraphics[width=240pt]{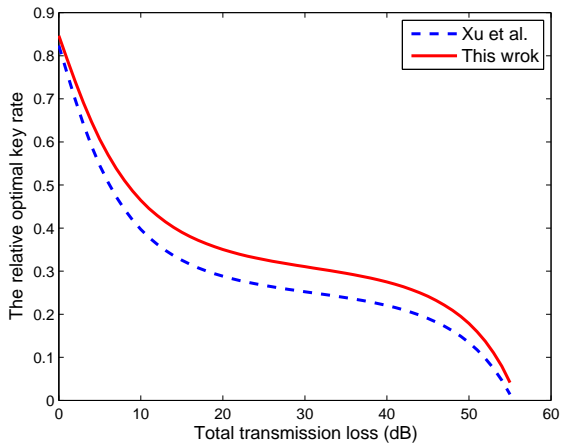}\\
  \caption{(Color online) The relative value between the optimal key rate obtained with different methods and the asymptotic limit of the infinite decoy-state method versus the total channel transmission loss using 3-intensity decoy state MDI-QKD. In the simulations, we assume that $\mu_v=\nu_v=0.01$, $\mu_d=\nu_d=0.1$, $\mu_s=\nu_s$ and the value of other parameters are presented in Table~\ref{tabPara}. The optimal key rate is just the value with maximizing the key rate with $\mu_s=\nu_s\in(\mu_d,1)$. }\label{rOptR}
\end{figure}

\begin{figure}
  \includegraphics[width=240pt]{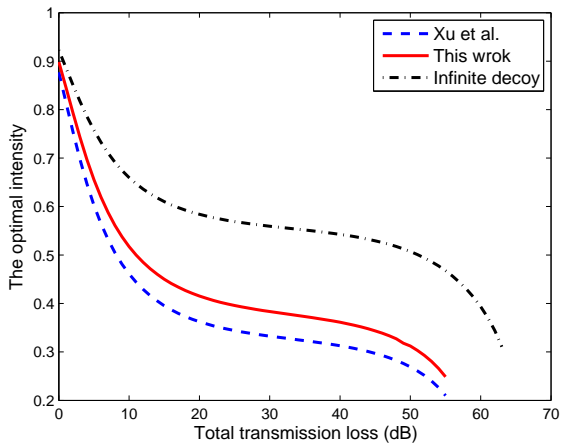}\\
  \caption{(Color online) The optimal intensities versus the total channel transmission loss using 3-intensity decoy state MDI-QKD. In the simulations, we assume that $\mu_v=\nu_v=0.01$, $\mu_d=\nu_d=0.1$, $\mu_s=\nu_s$ and the value of other parameters are presented in Table~\ref{tabPara}. The optimal densities of the signal-state pulses are obtained by maximizing the key rate.}\label{OptRmu}
\end{figure}

\begin{table}
\caption{\label{tabPara}List of experimental parameters used in numerical simulations: $e_0$ is the error rate of background, $e_d$ is the misalignment-error probability; $p_d$ is the dark count rate per detector; $f$ is the error correction inefficiency.}
\begin{ruledtabular}
\begin{tabular}{cccc}
  $e_0$ & $e_d$ & $p_d$ & $f$ \\
  \hline
  0.5 & 1.5\% & $3.0\times 10^{-6}$ & 1.16\\
\end{tabular}
\end{ruledtabular}
\end{table}

Now, we present some numerical simulations to comparing our results with the existing results~\cite{lopa,curty}. Below for simplicity, we suppose that Alice and Bob use the coherent-state sources.  Here, we denote Alice's sources $\{v_A,d_A,s_A\}$ by their intensities $\{\mu_v,\mu_d,\mu_s\}$ and Bob's sources $\{v_B,d_B,s_B\}$ by their intensities $\{\nu_v,\nu_d,\nu_s\}$ respectively.  The UTP locates in the middle of Alice and Bob, and the UTP's detectors are identical, i.e., they have the same dark count rate and detection efficiency, and their detection efficiency does not depend on the incoming signals. We shall estimate what values would be probably observed for the gains and error rates in the normal cases by the linear models as in~\cite{ind2,Wang2013,Wang201309}:
\begin{eqnarray*}
  |n\rangle\langle n| = \sum_{k=0}^n C_n^k \xi^k (1-\xi)^{n-k}|k\rangle\langle k|
\end{eqnarray*}
where $\xi^k$ is the transmittance for a distance from Alice to the UTB.  For fair comparison, we use the same parameter values used in~\cite{ind2,Wang2013,Wang201309} for our numerical evaluation, which follow the experiment reported in~\cite{UrsinNP2007}. For simplicity, we shall put the detection efficiency to the overall transmittance $\eta=\xi^2 \zeta$. We assume all detectors have the same detection efficiency $\zeta$ and dark count rate $p_d$. The values of these parameters are presented in Table~\ref{tabPara}. With this, the total gains $Y_{\mu_i,\nu_j}^{\omega},(\omega=X,Z)$ and error rates $Y_{\mu_i,\nu_j}^{\omega}E_{\mu_i,\nu_j}^{\omega},(\omega=X,Z)$ of Alice's intensity $\mu_i (i\in\{v,d,s\})$ and Bob's intensity $\nu_j (j\in\{v,d,s\})$ can be calculated. By using these values, we can estimate the lower bounds of yield $y_{11}^{Z}$ with Eq.(\ref{y11L123}) and Eq.(\ref{y11of14}). Also, we can estimate the upper bounds of error rate $e_{11}^{X}$ with Eq.(\ref{e11U}) where the lower bound of $y_{11}^{X}$ being estimated by Eq.(\ref{y11L123}) and Eq.(\ref{y11of14}). The estimated parameters of $y_{11}$ and $e_{11}$ are shown in Fig.\ref{ry11Lp1p5} and Fig.\ref{e11Up1p5}, respectively, which clearly shows that our methods are more tightly than the pre-existed results. In order to see more clearly, in Fig.\ref{ry11Lp1p5}, we plot the relative value of $y_{11}$ to the result obtained with the infinite decoy-state method. We can observe that our results are more close to the asymptotic limit of the infinite decoy-state method than the pre-existed results~\cite{lopa,curty}. Furthermore, with these parameters, we can estimate the final key rate $R$ of this protocol with Eq.(\ref{KeyRate}) which is shown in Fig.\ref{Rp1p5}. In these figures, the blue dashed line is obtained by Eq.(\ref{y11of14}) which is just the results presented in Refs.~\cite{lopa,curty}, the red solid line is obtained by the method presented in section 2 with Eq.(\ref{y11L123}), the black dash-dot line is the result obtained by the infinite decoy-state method. In the simulation, the densities used by Alice and Bob are assigned to $\mu_v=\nu_v=0.01$, $\mu_d=\nu_d=0.1$,  $\mu_s=\nu_s=0.5$.

Furthermore, if we fix the densities of the decoy-state pulses used by Alice and Bob, the final key rate will change with Alice and Bob taking different densities for their signal-state pulses. Then we can find out the optimal densities of their signal-state pulses with maximizing the final key rate. Here, we also take $\mu_v=\nu_v=0.01$, $\mu_d=\nu_d=0.1$ and assume that $\mu_s=\nu_s\in(\mu_d,1)$. In Fig.\ref{rOptR}, we plot the relative value of the optimal key rate to the result obtained with the infinite decoy-state method. We can observe that our result is better than the pre-existed results. The optimal densities with the optimal key rate versus the total channel transmission loss is given in Fig.\ref{OptRmu}.

\section{Conclusion}
We study the MDI-QKD in practice where the intensities of all  3 different states can be nonzero. Our result here is the most general result for the 3-intensity method: Setting the weakest state to be zero, we obtain the result in the special case where a vacuum state and two non-vacuum states are used\cite{Wang2013}. The result here is not limited to the coherent states. It applies to any states that satisfy Eq.(\ref{cond1}), e.g., the heralded states from the parametric down conversion. Our result is most efficient: It offers a more tightened bound for $y_{11}$ and therefore a higher key rate than the prior art\cite{lopa}, as has been strictly proven and also numerically demonstrated. The key rate can be further improved when it is combined with the method as shown in Ref.\cite{Wang201309}.

{\bf Acknowledgement:}
We acknowledge
the support from the 10000-Plan of Shandong province,
the National High-Tech Program of China Grants
No. 2011AA010800 and No. 2011AA010803 and NSFC
Grants No. 11174177 and No. 60725416.



\begin{thebibliography}{99}
\bibitem{BB84}
C.H.~Bennett and
G.~Brassard, in {\em Proc.\ of IEEE Int.\ Conf.\ on Computers,
Systems, and Signal Processing (IEEE, New York, 1984)},
pp.~175-179.
\bibitem{GRTZ02}
N.~Gisin, G.~Ribordy, W.~Tittel, and H.~Zbinden, Rev. Mod. Phys.
{\bf 74}, 145 (2002); N. Gisin and R. Thew, Nature Photonics, 1, 165
(2006); M.~Dusek, N.~L\"utkenhaus, M.~Hendrych, in {\em Progress in
Optics VVVX}, edited by E.~Wolf (Elsevier, 2006); V. Scarani, H.
Bechmann-Pasqunucci, N.J. Cerf, M. Dusek, N. L\"utkenhaus, and M
Peev, Rev. Mod. Phys. {\bf{81}}, 1301 (2009).
\bibitem{ILM}
H.~Inamori, N.~L\"utkenhaus, D.~Mayers, European Physical Journal D,
{\bf{41}}, 599 (2007), which appeared in the arXiv as quant-ph/0107017;
D.~Gottesman, H.K.~Lo, N.~L\"{u}tkenhaus, and J.~Preskill, Quantum
Inf. Comput. {\bf 4}, 325 (2004).

\bibitem{H03}
W.-Y.~Hwang, Phys. Rev. Lett. {\bf 91}, 057901 (2003).
\bibitem{wang05}
X.-B.~Wang, Phys. Rev. Lett. {\bf 94}, 230503 (2005).
\bibitem{LMC05}
H.-K.~Lo, X.~Ma, and K.~Chen, Phys. Rev. Lett. {\bf 94}, 230504
(2005).
\bibitem{AYKI}Y. Adachi, T. Yamamoto, M. Koashi, and N. Imoto, Phys. Rev.Lett.
{\bf 99}, 180503 (2007).
\bibitem{haya}M. Hayashi, Phys. Rev. A {\bf{74}}, 022307 (2006); ibid {\bf{76}},
012329 (2007).
\bibitem{peng}   D. Rosenberg {\em et al.},  Phys. Rev. Lett. {\bf{98}}, 010503
(2007);  T. Schmitt-Manderbach {\em et al.}, Phys. Rev. Lett.
{\bf{98}}, 010504 (2007); Cheng-Zhi Peng {\em et al.}
 Phys. Rev. Lett. {\bf{98}}, 010505 (2007); Z.-L. Yuan, A. W. Sharpe, and A. J. Shields,
Appl. Phys. Lett. {\bf{90}}, 011118 (2007); Y.~Zhao, B. Qi, X. Ma, H.-K.
Lo and L. Qian, Phys. Rev. Lett. {\bf 96}, 070502 (2006); Y. Zhao,
B. Qi, X. Ma, H.-K. Lo, and L. Qian, in Proceedings of IEEE
International Symposium on Information Theory, Seattle, 2006, pp.
2094--2098 (IEEE, New York).
\bibitem{wangyang} X.-B. Wang, C.-Z. Peng {\em et al.} Phys. Rev.
A {\bf{77}}, 042311 (2008);  J.-Z. Hu and X.-B. Wang, Phys. Rev. A, {\bf{82}}, 012331(2010).
\bibitem{rep}X.-B. Wang, T. Hiroshima, A. Tomita, and M. Hayashi, Physics Reports {\bf{448}}, 1(2007).
\bibitem{njp}X.-B. Wang, L. Yang, C.-Z. Peng and J.-W. Pan, New J. Phys. {\bf{11}}, 075006
(2009).
\bibitem{PNS1}
G.~Brassard, N.~L\"utkenhaus, T.~Mor, and
B.C.~Sanders, Phys. Rev. Lett. {\bf 85}, 1330 (2000);
N.~L\"utkenhaus, Phys. Rev. A {\bf 61}, 052304 (2000);
N.~L\"utkenhaus and M.~Jahma, New J. Phys. {\bf 4}, 44 (2002).
\bibitem{PNS}
B.~Huttner, N.~Imoto, N.~Gisin, and T.~Mor, Phys. Rev. A {\bf 51},
1863 (1995); H.P.~Yuen, Quantum Semiclassic. Opt. {\bf 8}, 939 (1996).
\bibitem{lyderson}L. Lyderson {\em et al}, Nature Photonics, {\bf{4}}, 686(2010); I. Gerhardt
{\em et al}, Nature Commu. {\bf{2}}, 349 (2011).
\bibitem{ind1}D. Mayers and A. C.-C. Yao, in Proceedings of the 39th
Annual Symposium on Foundations of Computer Science
(FOCS98) (IEEE Computer Society, Washington, DC,
1998), p. 503; A. Acin et al., Phys. Rev. Lett. {\bf{98}},
230501 (2007); Scarani V and Renner R Phys. Rev. Lett. {\bf{100}}, 302008 (2008);
Scarani V and Renner R 2008 3rd Workshop on Theory of Quantum Computation, Communication and
Cryptography (TQC 2008), (University of Tokyo, Tokyo 30 Jan¨C1 Feb 2008) See also arXiv:0806.0120
\bibitem{ind3}S.L. Braunstein and S. Pirandola, Phys. Rev. Lett. {\bf{108}}, 130502 (2012).
\bibitem{ind2}H.-K. Lo, M. Curty, and B. Qi, Phys. Rev. Lett., {\bf{108}},130503(2012), K. Tamaki et al, Phys. Rev. A, {\bf{85}}, 042307 (2012).
\bibitem{qing3}Qin-Wang abd Xiang-Bin Wang, arXiv:1305.6480.
\bibitem{wangPRA2013}
  Xiang-Bin Wang, Phys. Rev. A {\bf{87}}, 012320 (2013).
\bibitem{tittel1}Allison Rubenok, Joshua A. Slater, Philip Chan, Itzel Lucio-Martinez, Wolfgang Tittel, 1304.2463v1.
\bibitem{tittel2}P. Chan, J. A. Slater, I. Lucio-Martinez, A. Rubenok, W. Tittel, arxiv:1204.0738v1.
\bibitem{liuyang}Y. Liu et al, arXiv:1209.6178v1.
\bibitem{lopa}Feihu Xu, Marcos Curty, Bing Qi, Hoi-Kwong Lo , arXiv:1305.6965v1.
\bibitem{han} Chun Zhou, Wan-Su Bao, Wei Chen, Hong-Wei Li, Zhen-Qiang Yin, Yang Wang, Zheng-Fu Han, arXiv: 1308.3374v1.
\bibitem{curty}M. Curty et al, arXiv:1307.1081v1.
\bibitem{Wang2013}
  Z.-W. Yu, Y.-H. Zhou, X.-B. Wang, arXiv:1308.5677.
\bibitem{Wang201309}
  Z.-W. Yu, Y.-H. Zhou, X.-B. Wang, arXiv:1309.0471.
\bibitem{UrsinNP2007}
  R. Ursin, F. Tiefenbacher, T. Schmitt-Manderbach, H. Weier, T. Scheidl, M. Lindenthal, B. Blauensteiner, T. Jennewein, J. Perdigues, P.Trojek, B. \:{O}emer, M. F\:{u}erst, M. Meyenburg, J. Rarity, Z. Sodnik, C. Barbieri, H. Weinfurther, and A. Zeilinger, Nat. Phys. {\bf{3}}, 481 (2007).
\end{thebibliography}
\end{document}